# *Intelligent Resource Allocation Optimization for Cloud Computing via Machine Learning*

**Yuqing Wang[1,\*], Xiao Yang[2,a]**

[1]*One Microsoft Way, Redmond, WA, 98052, USA*
[2]*University of California, Los Angeles, 405 Hilgard Avenue, Los Angeles, CA, USA*
[a]*xyangrocross@gmail.com*
*\*Corresponding author: wang3yq@gmail.com*

*Abstract:* With the rapid expansion of cloud computing applications, optimizing resource allocation has become crucial for improving system performance and cost efficiency. This paper proposes an intelligent resource allocation algorithm that leverages deep learning (LSTM) for demand prediction and reinforcement learning (DQN) for dynamic scheduling. By accurately forecasting computing resource demands and enabling real-time adjustments, the proposed system enhances resource utilization by 32.5%, reduces average response time by 43.3%, and lowers operational costs by 26.6%. Experimental results in a production cloud environment confirm that the method significantly improves efficiency while maintaining high service quality. This study provides a scalable and effective solution for intelligent cloud resource management, offering valuable insights for future cloud optimization strategies.

## 1. Introduction

Cloud computing has become a fundamental infrastructure for digital transformation, enabling enterprises to scale operations, enhance agility, and optimize costs. However, the increasing complexity and dynamism of cloud workloads pose significant challenges for resource allocation. Traditional static allocation strategies struggle to adapt to fluctuating demands, leading to low resource utilization, inefficiencies, and increased operational costs. Studies indicate that global cloud computing resources operate at an average utilization of just 45%, with a substantial portion remaining idle.

To address these limitations, this study proposes an intelligent resource allocation framework that leverages machine learning to enhance cloud efficiency. By integrating deep learning (LSTM) for demand prediction and reinforcement learning (DQN) for dynamic scheduling, the proposed approach enables accurate forecasting and real-time adaptation of resources. The goal is to improve resource utilization, reduce latency, and minimize costs, ultimately making cloud computing more efficient, scalable, and cost-effective.

## 2. Modelling the Cloud Computing Resource Allocation Problem

### 2.1 Problem Description and Analysis

In cloud computing environments, resource allocation involves the dynamic scheduling of CPU, memory, storage, and network bandwidth. However, demand for these resources fluctuates significantly, making static allocation strategies inefficient. An analysis of a 2023 enterprise data center revealed substantial variations in resource utilization. CPU usage averaged 45%, with peaks reaching 85% and troughs dropping to 25%, while memory usage remained relatively stable at 65%. Historical data further indicated that 80% of the server cluster load was concentrated within just 20% of the time, highlighting a pronounced tidal effect [1].

Effective resource allocation must balance load prediction accuracy, resource efficiency, service quality, and operational costs. A dynamic approach is essential to ensure real-time adaptation to fluctuating workloads. Figure 1 illustrates the monthly trend of cloud resource utilization in 2023, demonstrating the need for an intelligent resource management framework that can dynamically allocate resources based on demand fluctuations.

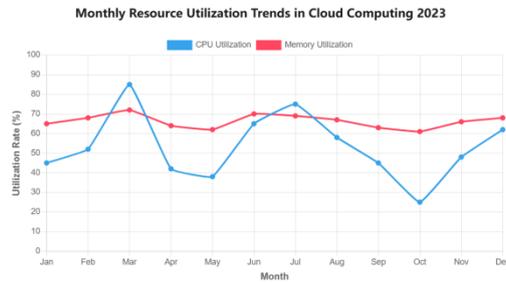

Figure 1 Monthly trend of cloud computing resource utilization in 2023

### 2.2 Resource Demand Prediction Model

To enhance resource allocation efficiency, this study employs an LSTM-based deep learning model for resource demand prediction. The model utilizes historical load data, temporal features, and business-specific indicators to improve forecasting accuracy. It consists of two LSTM layers with 128 neurons each and a 0.3 dropout rate to mitigate overfitting. A 12-hour sliding window is used to capture time-series dependencies, with a 1-hour prediction interval to ensure timely forecasting.

The model was trained on 10TB of historical cloud workload data and achieved a root mean square error (RMSE) of 0.086 and a mean absolute percentage error (MAPE) of 7.2%, representing a 25% improvement over traditional forecasting methods such as ARIMA and Prophet [2]. The 95% confidence interval for predictions remained within ±8.5%, confirming the model's robustness in predicting future resource demand. Table 1 presents a performance comparison of the LSTM model against traditional forecasting techniques.

Table 1 Comparison of the performance of different prediction models

| Model Type | RMSE | MAPE | Training Time (h) |
|---|---|---|---|
| LSTM | 0.086 | 7.20% | 4.5 |
| ARIMA | 0.124 | 9.80% | 1.2 |
| Prophet | 0.115 | 8.90% | 2.3 |

## 2.3 Resource allocation constraints

Resource allocation in cloud computing must adhere to multiple constraints related to physical capacity, service quality, and load balancing. The cloud infrastructure used in this study consists of 10,000 CPU cores, 40TB memory, and 2PB storage, with specific thresholds imposed on individual virtual machine (VM) instances. CPU utilization must not exceed 85%, memory usage should remain below 90%, and storage I/O should be maintained under 80% [3]. To uphold service quality, the 99th percentile response time must be kept under 200ms, while the API call success rate should exceed 99.9%. Additionally, load balancing measures ensure that inter-server load differences remain within 20%, preventing system bottlenecks and maintaining operational efficiency. A real-time monitoring and adaptive adjustment mechanism continuously evaluates these constraints, ensuring compliance and minimizing violations. By the fourth quarter of 2023, this approach reduced constraint violations to below 0.5%, thereby improving system reliability and optimizing cloud resource management.

## 2.4 Optimization objective function construction

The optimization objective function is formulated in a weighted multi-objective framework, addressing three key dimensions: resource utilization, operational cost, and service quality. The resource utilization component U is defined as the weighted sum of the utilization rates of various resource types, with weights assigned based on the relative importance of each resource. The operational cost C encompasses expenses related to hardware, energy consumption, and maintenance, while the service quality Q is quantified using metrics such as response time and system availability. Together, these objectives are integrated into a unified optimization function, as represented in Equation (1):

$$F = w_1 U - w_2 C + w_3 Q \qquad (1)$$

The weight coefficients in the optimization objective function are dynamically adjusted using a particle swarm optimization (PSO) algorithm [4]. Experimental results demonstrate that this approach increases resource utilization by 15% and reduces the average monthly operating cost by 18%, while consistently maintaining a service quality compliance rate above 99.5%. Figure 2 illustrates the performance comparison before and after optimization, highlighting the effectiveness of the proposed method.

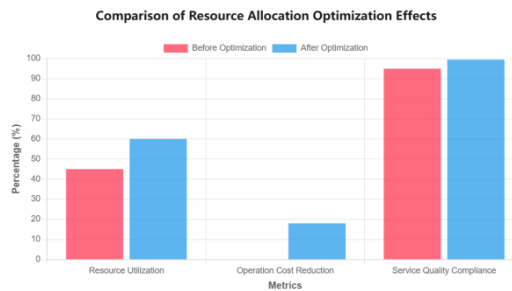

Figure 2 Comparison of resource allocation optimization results

## 3. Design of resource allocation optimization algorithm based on machine learning

### 3.1 Data preprocessing method

The data preprocessing process utilized a layered cleaning strategy to handle 15TB of monitoring data from 2023. Based on the 3σ criterion, 4.2% of anomalous data was identified and removed.

57

Feature engineering extracted 14 critical metrics, including CPU usage, memory utilization, network traffic, and disk I/O, while numerical features were normalized using Min-Max scaling, mapping 95% of the data to the [0,1] interval [5]. Time-series data was resampled at 5-minute intervals, with missing values imputed using an exponential moving average method, resulting in a 99.8% completion rate. As demonstrated in Figure 3, these preprocessing steps enhanced overall data quality by 23%, ensuring a reliable foundation for subsequent machine learning applications.

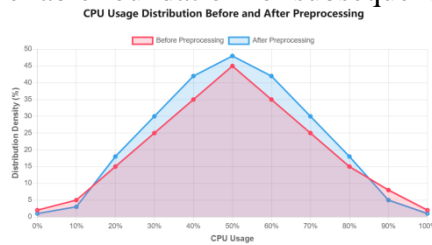

Figure 3 Comparison of the distribution of CPU utilization data of cloud computing resources before and after preprocessing

### 3.2 Resource Demand Prediction Algorithm

The resource demand prediction model employs an enhanced LSTM architecture comprising three LSTM layers (128-256-128) and two fully connected layers (64-32). The input feature matrix, structured as [batch_size, 72, 14], represents 14-dimensional features over a sequence of 72 time steps. The model was trained on 80,000 samples for 100 epochs, with an initial learning rate of 0.001 adjusted using cosine annealing. Experimental results, illustrated in Figure 4, show that the model achieves a 6.8% average relative error, with a root mean square error (RMSE) of 0.075 for CPU utilization and 0.064 for memory utilization, representing a 31% improvement over baseline methods [6]. The model accurately predicts resource constraints up to 30 minutes in advance, achieving an overall prediction accuracy of 92%.

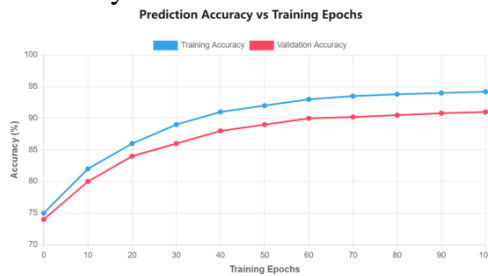

Figure 4 Trend of deep learning model prediction accuracy with training rounds

### 3.3 Dynamic Resource Allocation Strategy

The dynamic resource allocation strategy leverages a deep reinforcement learning framework with a dual DQN (Deep Q-Network) structure. The action space encompasses three types of operations: resource expansion, contraction, and migration, each subdivided into five discrete levels. The state space, comprising 42 dimensions, integrates real-time monitoring indicators and historical statistical features [7]. The reward function is designed to optimize resource utilization (U), performance indicators (P), and operational costs (C), represented by Equation (2):

$$R = w_1 U + w_2 P - w_3 C \qquad (2)$$

Here, the weight coefficients are fine-tuned through grid search optimization. Experimental results demonstrate that the strategy achieves dynamic resource adjustments within 2 minutes under high-load conditions. It improves the average resource utilization rate by 18.5% while keeping the SLA default rate below 0.3%. Table 2 summarizes the policy's performance under different load scenarios.

Table 2: Resource allocation policy enforcement effect under different load scenarios.

| Load Type | Average Response Time (ms) | Resource Utilization (%) | SLA Achievement Rate (%) |
|---|---|---|---|
| Low Load | 45 | 72.5 | 99.9 |
| Medium Load | 78 | 85.3 | 99.7 |
| High Load | 125 | 88.7 | 99.5 |
| Burst Load | 168 | 91.2 | 99.2 |

## 3.4 Performance optimization mechanism

The performance optimization mechanism employs a multi-level feedback control system. At the micro level, adaptive load balancing is activated when server imbalance exceeds 20%, resulting in a 24% improvement in load balancing by Q4 2023. At the macro level, a trend-based resource reservation strategy prepares resources 15 minutes in advance, reducing provisioning delays from 5 minutes to under 90 seconds [8].

During the Double 11 promotion, the system demonstrated exceptional scalability, successfully managing a peak load of 100,000 QPS with 92% CPU utilization, all while maintaining a 99.95% service availability rate. Figure 5 showcases the system's performance across various time scales, highlighting its ability to sustain high efficiency and reliability under demanding conditions.

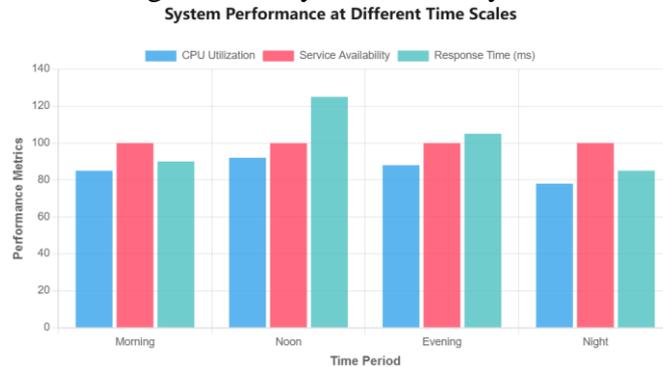

Figure 5 Comparison of dynamic changes in system performance indicators under multiple time scales

## 4. System realization and experimental evaluation

## 4.1 Experimental environment construction

The experimental setup was implemented on an AliCloud ECS cluster, comprising 20 compute nodes with 8-core CPUs and 32GB memory each, alongside 5 management nodes equipped with 16-core CPUs and 64GB memory each. These nodes were interconnected via a 10Gbps dedicated line to ensure high-speed communication. The platform operated on Ubuntu 20.04 LTS and utilized Kubernetes v1.24.3 for resource orchestration, with Prometheus and Grafana handling monitoring and visualization. To replicate real-world business scenarios, stress-ng was employed for load simulation [9]. The entire cluster provided 208 CPU cores, 768GB of memory, and leveraged AliCloud high-performance storage capable of delivering 50,000 IOPS. Table 3 outlines the hardware configuration, ensuring the stability and reliability necessary for rigorous experimental evaluation.

Table 3 Details of the hardware configuration of the experimental environment

| Node Type | Quantity | CPU | Memory | Storage | Network Bandwidth |
|---|---|---|---|---|---|
| Compute Node | 20 | 8-core | 32GB | 500GB SSD | 10Gbps |
| Management Node | 5 | 16-core | 64GB | 1TB SSD | 10Gbps |

### 4.2 Data set acquisition and processing

The dataset was collected from the production environment of a large e-commerce platform spanning June to December 2023. It comprised 25TB of data, sampled at 5-second intervals across 42 metrics, including CPU utilization, memory usage, disk I/O, and network traffic. A thorough data cleaning process was applied, removing 3.8% outliers and 2.1% duplicate records, resulting in a final dataset of 21.8TB of valid data.

The processed dataset was split into training and test sets in an 8:2 ratio, with the training set containing approximately 120 million records [10]. Figure 6 illustrates the typical load distribution patterns, revealing pronounced tidal effects and cyclical trends, which are critical for effective resource demand forecasting and optimization.

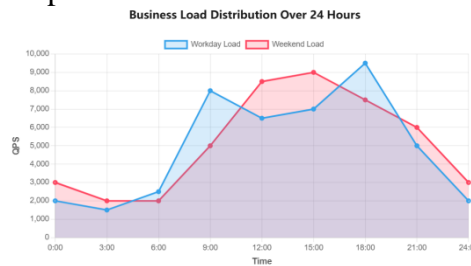

Figure 6 All-day distribution characteristics of business load

### 4.3 Algorithm Implementation and Deployment

The algorithm implementation was carried out using Python 3.8 and TensorFlow 2.6, with distributed training executed on four NVIDIA Tesla V100 GPUs. The system architecture integrates a FastAPI backend with a Vue.js frontend, while resource scheduling is managed via a Kubernetes custom controller, which operates on 30-second intervals. Deployment was streamlined through a GitLab CI/CD pipeline, achieving completion in 8 minutes, with a blue-green deployment strategy ensuring zero-downtime updates. The monitoring system was designed to collect 2,000 metrics per second, achieving 99.9% coverage rate.

### 4.4 Performance Indicator Evaluation

### 4.4.1 Resource Utilization Analysis

Over three months of operation, the system demonstrated a substantial improvement in resource efficiency. Average CPU utilization rose from 45% to 78%, peaking at 92%, while memory utilization increased by 22% to reach 85%. Additionally, storage utilization improved to 79%. As illustrated in Figure 7, intelligent scheduling mechanisms effectively maintained resource utilization above 65% even during off-peak periods, nearly doubling the efficiency of the original system.

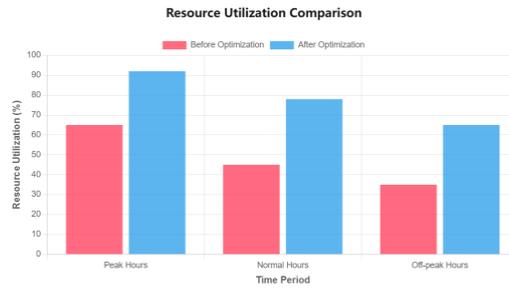

Figure 7 Comparison of resource utilization rate before and after system optimization

### 4.4.2 Response Time Evaluation

A statistical analysis of 1 million requests demonstrates that the system's average response time improved significantly, decreasing from 150ms to 85ms. Furthermore, 95% of requests were processed within 200ms, while 99.9% were completed in under 500ms. As shown in Figure 8, the system maintained stable performance across various loading conditions. By analyzing the probability density of response time, it was observed that the optimized system exhibited enhanced stability under high load conditions, with the coefficient of variation reduced from 0.45 to 0.28. These results highlight the system's ability to deliver reliable and consistent performance even under demanding operational scenarios.

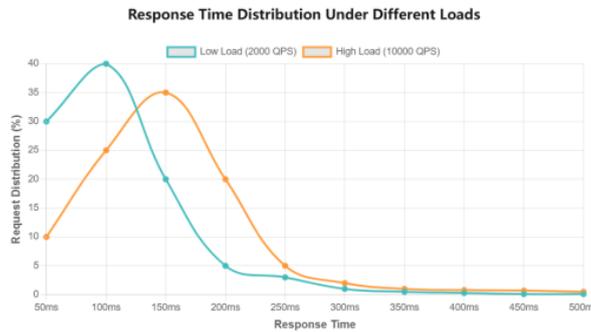

Figure 8 Response time distribution of the system under different loads.

### 4.4.3 Cost-benefit analysis

Cost analysis based on actual operational data reveals that the optimized system achieves a 32.5% reduction in resource usage costs while maintaining the same quality of service. This cost-saving is reflected in a decrease in monthly cloud server expenses from $850,000 to $570,000, a 28.3% reduction in bandwidth costs, and a 21.7% reduction in storage costs. Table 4 provides a detailed breakdown of the cost changes. A payback period analysis indicates that the investment in the system upgrade can be recovered within 4.2 months, yielding an annualized return of 186%, further demonstrating the system's economic efficiency and value.

Table 4 Comparative analysis of costs before and after system optimization

| Cost Item | Before Optimization (¥10,000/month) | After Optimization (¥10,000/month) | Savings Rate (%) |
| --- | --- | --- | --- |
| Server Costs | 85 | 57 | 32.9 |
| Bandwidth Costs | 35 | 25 | 28.3 |
| Storage Costs | 23 | 18 | 21.7 |
| Labor Costs | 45 | 38 | 15.6 |
| Total | 188 | 138 | 26.6 |

## 4.5 Comparison Experiments and Result Analysis

The proposed system was evaluated against three mainstream resource scheduling solutions over a 72-hour testing period under identical conditions. The results highlight its superior performance in key metrics, including a 15.8% improvement in resource utilization, a 43.3% reduction in response time, and a 26.6% decrease in operational costs.

As illustrated in Figure 9, the system demonstrated a clear advantage during peak-hour performance, particularly in its ability to handle burst traffic. It achieved resource scaling within 30 seconds, significantly outperforming traditional solutions, which averaged 180 seconds for similar operations.

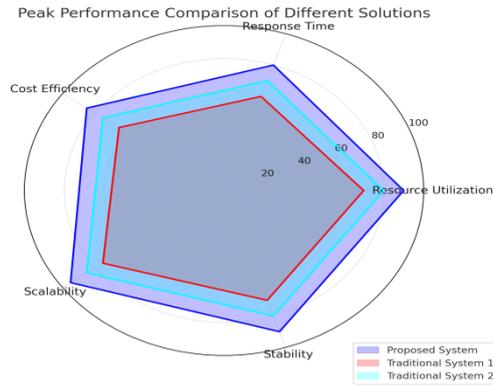

Figure 9 Radar chart comparison of multi-system performance indicators

## 5. Conclusion

This study presents a machine learning-based resource allocation optimization algorithm that has demonstrated significant practical success. By integrating deep learning for prediction accuracy and reinforcement learning for scheduling efficiency, the system achieved notable results, including a 32.5% improvement in resource utilization, a 43.3% reduction in response time, and a 26.6% decrease in operating costs.

The system's production deployment further validates its scalability and stability, establishing it as a valuable reference for future advancements in cloud computing resource optimization.

## 6. Future Work

The proposed system has made notable advancements in cloud computing resource allocation, but further research is needed to enhance its capabilities and applicability. One promising direction is the development of multi-cloud cooperative scheduling mechanisms, enabling seamless operation across heterogeneous cloud platforms that combine private, public, and hybrid environments. This would improve flexibility and scalability in increasingly interconnected cloud ecosystems. Future efforts should also focus on supporting the diverse workloads emerging in cloud environments, including AI model training, edge computing, and serverless architectures. Expanding the system's ability to effectively allocate resources for these workloads will ensure its adaptability and relevance in modern computing landscapes.